\documentclass{article}

\usepackage{arxiv}

\usepackage{algorithmic}
\usepackage{textcomp}
\usepackage{xcolor}
\usepackage{hyperref}
\usepackage{listings}
\usepackage{framed}
\usepackage{amsfonts}
\usepackage{multirow}
\usepackage{graphicx} 
\usepackage[frozencache,cachedir=.]{minted}

\AtBeginDocument{%
  }

\title{Hardcaml: An OCaml Hardware Domain-Specific Language for Efficient and Robust Design}

\date{20 December 2023}

\author{
  Andy Ray, Benjamin Devlin, Fu Yong Quah, Rahul Yesantharao \\
  Jane Street \\
  \texttt{\{aray, bdevlin, fquah, rayesantharao\}@janestreet.com}
}

\renewcommand{\headeright}{}
\renewcommand{\undertitle}{}
\renewcommand{\shorttitle}{Hardcaml}

\begin{document}

\maketitle

% \author{Andy Ray}
% \email{aray@janestreet.com}
% \affiliation{%
%   \institution{Jane Street}
%   \city{London}
%   \country{UK}
% }
% \orcid{0009-0000-9512-2606}
% 
% \author{Benjamin Devlin}
% \email{bdevlin@janestreet.com}
% \affiliation{%
%   \institution{Jane Street}
%   \city{New York City}
%   \country{USA}
% }
% \orcid{0009-0008-2823-549X}
% 
% \author{Fu Yong Quah}
% \email{fquah@janestreet.com}
% \affiliation{%
%   \institution{Jane Street}
%   \city{London}
%   \country{UK}
% }
% \orcid{0009-0002-0091-5481}
% 
% \author{Rahul Yesantharao}
% \email{ryesantharao@janestreet.com}
% \affiliation{%
%   \institution{Jane Street}
%   \city{New York City}
%   \country{USA}
% }
% \orcid{0009-0009-7786-0969}

\begin{abstract}

This paper introduces Hardcaml, an embedded hardware design domain specific language (DSL)
implemented in the OCaml programming language. Unlike high level synthesis (HLS), Hardcaml
allows for low level control of the underlying hardware for maximum productivity, while
abstracting away many of the tedious aspects of traditional hardware definition languages (HDLs) such as
Verilog or VHDL.
The richness of OCaml's type system combined with Hardcaml's fast circuit elaboration
checks reduces the chance of user-introduced bugs and erroneous connections with features
like custom type defining, type-safe parameterized modules and elaboration-time bit-width
inference and validation.
Hardcaml tooling emphasizes fast feedback through simulation, testing, and verification.
It includes both a native OCaml cycle-accurate and an event-driven simulator.
Unit tests can live in the source code and include digital ASCII waveforms representing
the simulator's output. Hardcaml also provides tools for SAT proving and formal verification.
Hardcaml is industrially proven, and has been used at Jane Street internally for many
large FPGA designs.
As a case study we highlight several aspects of our recent Hardcaml submission to the 2022 ZPrize
cryptography competition which won 1st place in the FPGA track.

\end{abstract}


%%
%% Keywords. The author(s) should pick words that accurately describe
%% the work being presented. Separate the keywords with commas.
% \keywords{CAD, RTL, Hardware Domain-Specific Languages, Efficiency, Robustness, FPGA}

%\received{13 October 2023}
%\received[revised]{12 March 2009}
%\received[accepted]{5 June 2009}

\maketitle

\section{Introduction}

(System) Verilog and VHDL are currently the most commonly used hardware definition
languages (HDLs) used in industry. Despite being expressive enough to create hardware
circuits, these languages have significant differences from software languages; they have
correspondingly not benefited from advances in programming language design. These advances
improve productivity and flexibility, and are also crucial to hardware design as systems
become much more large and complex. In modern hardware development, the bulk of a hardware
designer's time is spent in modeling, simulation, and testing~\cite{testingtime}.

A large number of existing domain-specific languages (DSLs)~\cite{dsl, dsl-survey} aim at
solving the various shortcomings of Verilog and VHDL. Each of these provide different
levels of abstraction and control of underlying hardware primitives, guarantees of safety,
and additional features or meta-programming. The advantages of using such DSLs have been
shown empirically. An example RISC-V core design that required 1000 lines of hand-written
Verilog required only 586 lines of Chisel~\cite{chisel-vs-verilog}, and a similar design
required 637 lines of Hardcaml~\cite{hardcaml-risk5}. Reducing the number of lines of code
a designer has to write increases productivity and decreases the chance for bugs. There is
also a great interest in translating traditional RTL designs into the domain of these
DSLs~\cite{verilog-to-chisel}.

Existing DSLs can be divided into two main categories in terms of their abstraction from
the underlying RTL-- compiler style DSLs and full control DSLs. Compiler style DSLs allow
the designer to express a circuit in terms of dataflow or a higher level language such as
C++, which the DSL framework compiles to RTL. This increases productivity in hardware
design by fast iteration over potentially large design spaces containing many different
parameters, at the cost of preventing fine clock cycle-level control and optimization.
These are often referred to as high level synthesis (HLS) tools, and can be stand-alone
tools such as Vivado HLS, PyLog~\cite{pylog}, or wrappers around existing tool chains and
general-purpose compiler infrastructure~\cite{hlsrecursion,scalehls} such as
MLIR~\cite{mlir}. For example Bluespec~\cite{bluespec,bluespec2} allows design using lists
of declarative rules, which each fire automatically with handshaking to control data flow.

Full control DSLs allow writing hardware that can directly instantiate hardware primitives
such as registers, memories, multiplexors, and so on. They allow the designer to optimize
a circuit down to clock-cycle latencies, while offering improvements on top of traditional
HDL languages via software language paramertization and meta-programming. Hardcaml is a
full control DSL. Other full control DSLs are Filament~\cite{filament}, which while only
currently supporting statically scheduled pipelines, allows for the safe reusability of
modules through \textit{timeline types} that encode latency and throughput.
Chisel~\cite{chisel}, Clash~\cite{clash}, SpinalHDL~\cite{spinalhdl}, MyHDL~\cite{myhdl}
provide similar parametrization to Hardcaml, but each has a slightly different toolkit it
exposes to the designer.

%% understand this comment:I usually see a non-breaking space before citations like `[10]`.

OCaml modules and functors provide flexible circuit parameterization in Hardcaml. Both the
type system and an extensive testing and verification toolkit ensure safety and
correctness. Circuit elaboration is fast and happens during both RTL generation and
simulation, completely inside OCaml. During elaboration Hardcaml ensures there are no
floating wires, no multiple drivers, and when signals do connect that their widths match
exactly. The combination of these allow for RTL design that avoids common place bugs,
allowing an engineer to spend more time on the enjoyable and valuable areas of hardware
design.

Both a built-in cycle-accurate and event-driven simulator allows for unit level
tests alongside the Hardcaml source code, which can optionally print digital
ASCII waveforms. These tests provide fast feedback on designs and help catch
future bugs. Hardcaml can generate hardware design from abstraction models
written in OCaml, simulation expect-tests with waveforms, interactive waveform
viewers, synthesis reports, and finally allow production-ready Verilog and/or
VHDL.

The development flow for hardware development in Hardcaml is very similar to
software development. The availability of Hardcaml simulators and various
tooling, along with the existing OCaml Dune build system~\cite{dune} and
availability of various OCaml libraries, draws testing and simulation into the
ordinary software development life cycle. The comprehensive testing
capabilities, all within a single programming language, allow for rapid
hardware development with a fast feedback loop in simulation.

While existing DSLs address different aspects of safety and productivity,
Hardcaml does this while providing the hardware engineer with a more contained and complete solution. Other
DSLs do not include as comprehensive testing capabilities such as unit test waveforms or
built-in simulators. Each DSL is implemented inside a higher level software language,
which can ultimately define its capabilities and limitations. OCaml is a strongly
typed and modern functional programming language, and Hardcaml is entirely embedded inside
OCaml without the need for any secondary abstraction layer. This allows Hardcaml to be
lightweight and makes it easier to develop improvements or debug issues directly in the
source code. Hardcaml is an actively maintained open-source~\cite{hardcaml-github} project
developed at Jane Street, where it's the primary system used for creating production FPGA
designs.

\section{Hardcaml Overview}

Hardcaml is a DSL embedded in OCaml that improves reliability when writing HDL, while
remaining light weight and easy to use. OCaml is a multi-paradigm (functional,
imperative, object-oriented) high level programming language with an expressive type
system and type inference. Hardcaml is a collection of OCaml libraries that designers can
use to express circuits with the same amount of control as Verilog or VHDL, but with the
advantages of a higher-level programming language. It cuts out a lot of the mundane
error-prone work when connecting modules and parameterizing circuits, as will be
demonstrated. A strong built-in linter prevents bugs such as signals of different widths
driving each other, and detects floating and multiple drivers.

The typical workflow of designing with Hardcaml is as follows:

\begin{enumerate}
\item Write the circuit implementation using Hardcaml libraries.
\item Compose testbenches, which are run using the built-in Hardcaml simulator.
\item Debug using the built-in terminal-based waveform viewer and expect-tests.
\item Generate Verilog from OCaml using the built-in Hardcaml RTL generator.
\item Run generated Verilog through standard vendor tooling (Vivado, etc.).
\end{enumerate}

Steps 1-3 can leverage existing OCaml tooling such as
the Dune build system~\cite{dune} for rapid development.

\subsection{Datatypes}

The core type in a Hardcaml circuit is the \verb|Signal.t| type. Various functions
and operators that take \verb|Signal.t| as parameters represent hardware primitives.
Table \ref{table:signal_primitives} shows a few examples of these.

\begin{table}[!ht]
  \caption{Example Hardware Primitives}
{
    \begin{tabular}{|l|l|}
      \hline
      \verb|+:|, \verb|+:.| & Addition operator \\
      \hline
      \verb|-:|, \verb|-:.| & Subtraction operator \\
      \hline
      \verb|&:| & Bitwise AND operator \\
      \hline
      \verb+|:+ & Bitwise OR operator \\
      \hline
      \verb|==:|, \verb|==:.| & Equality operator \\
      \hline
      \verb|reg, reg_spec| & Registers and clock/clear specification \\
      \hline
      \verb|mux2|, \verb|mux| & Multiplexors \\
      \hline
    \end{tabular}}
    \label{table:signal_primitives}
\end{table}

All operators contain a \verb|:| to distinguish them between regular OCaml operators on
integers, as OCaml does not support operator overloading. There are multiple variants of
arithmetic operators (eg: \verb|+:| and \verb|+:.|). The \verb|+:| and \verb|+:.|
operators both represent addition. \verb|+:| operates on two \verb|Signal.t|s that must
match bit-width, while \verb|+:.| operates on a \verb|Signal.t| and integer that has its
bit-width inferred. \verb|reg_specs| specify clock and clear signals for sequential logic
such as \verb|regs|. \verb|mux| is a general multiplexor that takes a list of arguments,
and \verb|mux2| is a convenient helper function to write a simple if-then-else multiplexor
with two cases.

The programmer can create a hardware circuit by calling functions on a value of type
\verb|Signal.t|, and then storing the returned value (which is also a \verb|Signal.t|). For example,
the following function creates a counter that counts to \verb|count_to| and then wraps back to
zero.\verb|~| is used to pass a named argument to a function, and
\verb|()| is sometimes necessary to terminate a sequence of function arguments in
OCaml.

\begin{minted}{ocaml}
let counter ~clock ~count_to =
  let spec = Reg_spec.create ~clock () in
  let width = Int.ceil_log2 (count_to + 1) in
  let x = wire width in
  x <== reg spec (mux2 (x ==:. count_to) (zero width) (x +:. 1));
  x
\end{minted}

The code listing above constructs a counter which behaves similarly to the
Verilog code snippet below.

\begin{minted}{verilog}
module counter #(
  parameter COUNT_TO = 100,
  parameter CTR_WIDTH = $clog2(COUNT_TO + 1)
) (input clock,
   output reg [CTR_WIDTH-1:0] x);

  always @ (posedge clock) begin
    x <= x == COUNT_TO ? 0 : x + 1;
  end
endmodule
\end{minted}

The advantage of using Hardcaml here is it will prevent the counter output connecting to
something that isn't wide enough, which if happened in Verilog would silently drop the
unused bits.

\subsection{Defining New Types}

Hardcaml can represent each logical idea in its own type, defined by the programmer.
OCaml's type system then statically guarantees that values of a given type cannot be used
in ways other than those defined valid for that type. This enables both abstraction and
eliminates entire classes of potential errors. Note that these types are parameterized
with \verb|'a|, later to be replaced with \verb|Signal.t|s. For example, the following
Hardcaml snippet defines a type called \verb|Rectangle.t|, which has a 10-bit length and a
6-bit width. \verb|Signal.t Rectangle.t| is similar to a \verb|Rectangle<Signal>|
in Java or C++. A parameterized type argument allows us to use \verb|Rectangle.t| in
different contexts, like tests, debugging, and HDL generation.

\begin{minted}{ocaml}
module Rectangle = struct
  type 'a t =
    { length : 'a [@bits 10]
    ; width : 'a  [@bits 6]
    }
  [@@deriving hardcaml]
end
\end{minted}

You can then write a function that takes a \verb|Rectangle.t| as an argument. This snippet
would generate a circuit that outputs the area of a given rectangle.

\begin{minted}{ocaml}
let calculate_area
      ~clock
      (rectangle : Signal.t Rectangle.t) =
  let spec = Reg_spec.create ~clock () in
  reg spec (rectangle.length *: rectangle.width)
\end{minted}

\verb|[@@deriving hardcaml]| instructs the Hardcaml preprocessor to generate utility
functions working with the \verb|Rectangle.t| type. The derived \verb|Rectangle.map|
function applies a function to every field. For example, this function returns
a rectangle where both the length and width are increased by a certain size.

\begin{minted}{ocaml}
let extend_length_and_width ~by rectangle =
  Rectangle.map rectangle ~f:(fun x -> x +:. by)
\end{minted}

Another derived function is \break\verb|Rectangle.Of_signal.priority_select|, which
performs a combinational priority select on the custom rectangle type. For example, this
function takes a list of rectangles and outputs the first rectangle whose width is greater
than \verb|threshold|. It uses \verb|List.map| to convert the \verb|Rectangle.t| inputs to \verb|With_valid.t|
objects. A \verb|With_valid.t| object contains an input rectangle (\verb|value|) and a single bit
indicating whether the priority select should accept the input (\verb|valid|).

\begin{minted}{ocaml}
let choose_rectangle ~threshold rectangles =
  List.map rectangles ~f:(fun rectangle ->
    { With_valid.valid =
      rectangle.width >:. threshold
    ; value = rectangle
    }
  |> Rectangle.Of_signal.priority_select
\end{minted}

Custom types can be nested arbitrarily to create multiple layers of logical
separation. For example:

\begin{minted}{ocaml}
module Shape_collection = struct
  type 'a t =
    { rectangle1 : 'a Rectangle.t
    ; rectangle2 : 'a Rectangle.t
    }
  [@@deriving hardcaml]
end
\end{minted}

%Comment from Pranjal:
%I'm not sold on the importance of this section about scalar types. You need to introduce signatures and functors well before they're explained, and the fact that you can create scalars isn't (IMO) that interesting. The interesting part here is showing off the type safety. So I think instead, you could have an example that's like `Point.t` or `Cylinder.t` that also has two fields for $x$ and $y$ or $radius$ and $height$ (make them the same size as `Rectangle.t`), and then demonstrate that you can't accidentally try and get the area of a `Point.t`.

Another abstract type that one would define is a scalar-type. These are bit-vectors of a
particular width that have logical meaning. For example, we can define a 32-bit price in
US dollars. It's important that we don't use values of this type when we're expecting, for
example, Euros. We then can define some custom functions that operate on this type, namely \verb|is_gte_zero|,
\verb|min|, and \verb|max|. Including the module \verb| Types.Scalar| creates a type \verb|Price_in_usd.t| and common functions on that type.

\begin{minted}{ocaml}
module Price_in_usd : sig
  include Types.Scalar.S

  val is_gte_zero : Signal.t t -> Signal.t
  val min : Signal.t t -> Signal.t t -> Signal.t
  val max : Signal.t t -> Signal.t t -> Signal.t
end = struct
  include Types.Scalar.Make(struct
    let port_name = "price_in_usd"
    let port_width = 32
  end)

  let is_gte_zero t = t >=:. 0
  let min a b = mux2 (a <: b) a b
  let max a b = mux2 (a >: b) a b
end
\end{minted}

Functions can be defined to explicitly expect a value of type \verb|Price_in_usd.t|,
rather than an arbitrary value of type \verb|Signal.t|.

\begin{minted}{ocaml}
let cap_price_at_zero (x : _ Price_in_usd.t) =
  mux2 (Price_in_usd.is_gte_zero x)
    x
    (Price_in_usd.Of_signal.of_int 0)
\end{minted}

The OCaml compiler will permit the following.

\begin{minted}{ocaml}
cap_price_at_zero (Price_in_usd.Of_signal.of_int 42)
\end{minted}

On the other hand, the OCaml compiler will emit a compile-time error
due to a type mismatch in the following code snippet. This is an important
safety feature of Hardcaml as it prevents the programmer from passing values of
unexpected types into functions.

\begin{minted}{ocaml}
cap_price_at_zero (Signal.of_int ~width:32 42)
\end{minted}

Here is another example of a compile-time error due to a type mismatch as the
programmer uses the type \verb|Signal.t Rectangle.t| instead of
\verb|Signal.t Price_in_usd.t|, producing the error
\begin{minted}{ocaml}
let rectangle =
  { Rectangle.
    length = Signal.of_int ~width:10 1
  ; width  = Signal.of_int ~width:6  2
  }
in
cap_price_at_zero rectangle
(* Error: This expression has type Signal.t Rectangle.t but an expression was
expected of type Signal.t Price_in_usd.t *)
\end{minted}

These scalar types can also be nested within interfaces.

\begin{minted}{ocaml}
module Min_max = struct
  type 'a t =
    { min :  'a Price_in_usd.t
    ; max : 'a Price_in_usd.t
    }
  [@@deriving hardcaml]
end

let compute_min_and_max a b =
  { Min_max.
    min = Price_in_usd.min a b
  ; max = Price_in_usd.max a b
  }
\end{minted}

\subsection{Hierarchical Modules}

A Hardcaml circuit module can be structured similar to a Verilog circuit module.
Using the \verb|hierarchy| library inside Hardcaml, a global circuit
hierarchy can be constructed. This produces simulation waveforms with
hierarchical signals which increases debugging productivity, and generates
RTL with hierarchical modules, which works better with existing
Verilog and VHDL-based vendor tooling.

To define a Hardcaml circuit module, the programmer defines \verb|I| and \verb|O|
modules, corresponding to the input and output interfaces. Then, the programmer
defines a \verb|create| function that when given \verb|Signal.t I.t|, returns
\verb|Signal.t O.t|.
For example:

\begin{minted}{ocaml}
module I = struct
  type 'a t =
    { clock  : 'a
    ; x : 'a [@bits 10]
    }
  [@@deriving hardcaml]
end

module O = struct
  type 'a t =
    { y : 'a [@bits 20]
    }
  [@@deriving hardcaml]
end

let create (i : Signal.t I.t) =
  let spec = Reg_spec.create ~clock:i.clock () in
  { O.y = reg spec (i.x *: i.x) }
\end{minted}

\subsection{Bit-width Inference and Post Processing Validation}

Hardcaml handles bit-width strictly. The circuit construction step validates bit-widths
for all operations before running simulations or generating Verilog. This prevents many
errors at the compilation stage, increasing the speed of the designer's feedback loop. For
example, the following program will fail due to a width mismatch:

\begin{minted}{ocaml}
let not_allowed a b = uresize a 32 +: uresize b 16
\end{minted}

Hardcaml can also  infer the bit-width for most operations, so users
don't need to specify the width for most internal wires. In the case above we
specifically resized our wires to 32-bits and 16-bits.

For example, Hardcaml will infer the width of \verb|mult_result| as \break
\verb|width a + width b|
and the width of \verb|eq_result| as \verb|1|.

\begin{minted}{ocaml}
let foo a b =
  let mult_result = a *: b in
  let eq_result = a ==: b in
  ...
\end{minted}

During circuit elaboration, Hardcaml also performs other validation, ensuring that:
\begin{itemize}
\item Module instantiations have appropriate port widths.
\item All wires have exactly one driver.
\item Operation bit-widths match.
\end{itemize}
These errors are caught early on before simulation is run or RTL
is generated.

\subsection{The Always DSL}

Hardcaml includes the Always DSL, which can describe circuits in the same style as a
Verilog always block. Signals are now inside guarded assignments, either as asynchronous
wires or synchronous regs. The Always DSL does not support blocking assignments. Below we
show an example of an accumulating sum state machine written in this style. This module
will add and accumulate all inputs provided as long as any input is non-zero.

First we define a clock and clear module \verb|Control| which will ensure other modules
types cannot drive this circuit. This is then combined with our data \verb|inputs| which is an
OCaml list of signals, each with a bit-width of 4. Our output is a single signal of
bit-width 4.

\begin{minted}{ocaml}
module Adder = struct
  module Control = struct
    type 'a t =
      { clock : 'a
      ; clear : 'a
      }
    [@@deriving hardcaml]
  end
  module I = struct
    type 'a t =
      { control : 'a Control.t
      ; inputs : 'a list [@bits 4] [@length 2]
      }
    [@@deriving hardcaml]
  end
  module O = struct
    type 'a t =
      { output : 'a [@bits 4]
      }
    [@@deriving hardcaml]
  end
\end{minted}

We then define the the states our circuit can be in with the module \verb|State|, which
are \verb|Idle| and \verb|Adding|. We use the preprocessor to automatically derive
functions that let Hardcaml internally compare and enumerate in order to implement the state machine.

\begin{minted}{ocaml}
  module State = struct
    type t =
      | Idle
      | Adding
    [@@deriving enumerate, compare]
  end
\end{minted}

The Always DSL \verb|create| function takes this module as its input and returns
\verb|sm|, the control object of our state machine.

Inside the create function's \verb|control| input, we have expanded out the
\verb|clock| and \verb|clear| inputs, which means the named arguments for
\verb|Reg_spec.create| match and do not need to be specified, reducing the chance of a
connection typo.

\begin{minted}{ocaml}
  let create
      ({ control = { clock; clear }
       ; inputs
       } : _ I.t) =
    let spec = Reg_spec.create ~clear ~clock () in
    let sm =
      State_machine.create
        (module State)
        spec
    in
 ...
 \end{minted}

We then declare the output \verb|O.t| as an \verb|always reg|-- this means we can use the
Always DSL logic assignment operator (\verb|<--|) to assign values to our output from
within our state machine. For each state, we add logic for transitioning to new states and
assigning output values. We use the Hardcaml \verb|tree| and \verb|reduce| functions to
construct a log2 tree that will sum elements. This demonstrates how optimized data
structures can be used with low effort on the designers part. The preprocessor generates a
function \verb|O.Of_always.value| that converts our \verb|always reg| to a \verb|Signal.t O.t|.

 \begin{minted}{ocaml}
    let o = O.Of_always.reg spec  in
    compile [
      sm.switch [
        Idle, [
          when_ (concat_lsb inputs <>:. 0) [
            sm.set_next Adding
          ]
        ];
        Adding, [
          if_ (concat_lsb inputs ==:. 0) [
            sm.set_next Idle
          ] [
            o.output <-- (
              o.output.value +: tree ~arity:2 inputs ~f:(reduce ~f:( +: ))
            );
          ]
        ]
      ]
    ];
    O.Of_always.value o
end
\end{minted}

The Always DSL allows for further abstraction by using OCaml
functions. This also can provide implicit documentation via function
names in complicated state machines.

\begin{minted}{ocaml}
let decr_with_floor x =
  x <-- mux2 (x ==:. 0) x.value (x.value -:. 1);

let create =
  let foo = Var.reg spec ~width:8 in
  let bar = Var.reg spec ~width:8 in
  Always.(compile [
    decr_with_floor foo;
    decr_with_floor bar
  ]);
\end{minted}

\subsection{Module-Level Parameterization with Functors}

In OCaml a functor is a module that takes one or more modules as a parameter, and returns
a new module. These allow us to parameterize Hardcaml circuits, similar to Verilog's parameters. The
programmer first defines a config argument for the parameterized module. For example:

\begin{minted}{ocaml}
module type Config = sig
  val metadata_bits : int

  module Data : Hardcaml.Interface.S
end
\end{minted}

Then, they can define the corresponding circuit and module using fields
(\verb|metadata_bits|) or datatypes (\verb|Data.t|) from the functor argument.
The module \verb|Make| below is an OCaml functor.

\begin{minted}{ocaml}
module Make(X : Config) = struct
  module I = struct
    type 'a t =
      { clock : 'a
      ; metadata : 'a [@bits X.metadata_bits]
      ; data : 'a X.Data.t
      }
    [@@deriving hardcaml]
  end
end
\end{minted}

The functor can then be instantiated with an argument of the appropriate type
signature. For example:

\begin{minted}{ocaml}
module Foo = struct
  type 'a t =
    { bar : 'a
    ; baz : 'a
    }
  [@@deriving hardcaml]
end

module A = Make(struct
  let metadata_bits = 42

  module Data = Foo
end)
\end{minted}

This is similar to parameterized circuits in Verilog. The key advantage
to this functor-style is that the parameters themselves can be arbitrary arguments, such
as lists, integers, arrays or even custom types (\verb|Data.t| in the above case).
This allows for high expressiveness in parameter descriptions. When instantiating
the \verb|Make| functor, the OCaml compiler will ensure that the provided
argument value matches its expected type.

Functors, as demonstrated above, are a powerful feature in OCaml that allows
modularity and parametrization. This demonstrates an advantage of Hardcaml
being embedded in a programming language: it allows the programmer to
leverage existing software-level features for hardware design.

\subsection{Example Circuit Descriptions}

Below are some examples of circuit description that demonstrate Hardcaml's
expressiveness.

\subsubsection{Tree Adder}

Recursive circuits in Hardcaml come very naturally. The following circuit decomposes
operands into a tree and performs an arbitrary associative operation \verb|op| on them:

\begin{minted}{ocaml}
let rec tree_op ~op xs =
  List.chunks_of ~length:2 xs
  |> List.map ~f:(fun chunk ->
    List.reduce_exn ~f:op chunk)
  |> tree_op ~op
\end{minted}

The \verb|tree_op| can be used to construct various a tree of associatative
operators, such as a tree adder or a tree AND.

\begin{minted}{ocaml}
let tree_adder = tree_op ~op:(+:)
let tree_and   = tree_op ~op:(&:)
\end{minted}

Note that in practice, tree-like constructs can be expressed using the
\verb|tree| function. The example above serves to illustrate a recursive
example.

\subsubsection{Constructing a Linear Feedback Shift Register}

A Linear Feedback Shift Register (LFSR) is a shift register whose input bit is
the XOR of 2 of more bits (taps) of it's previous state. This can be
expressed in Hardcaml as follows:

\begin{minted}{ocaml}
let xor_lfsr ~width ~clock taps =
  let spec = Reg_spec.create ~clock () in
  reg_fb ~width spec ~f:(fun x ->
    let input_bit =
      List.map taps ~f:(fun i -> x.:(i))
      |> reduce ~f:(^:)
    in
    lsbs x @: input_bit)

let x = xor_lsfr ~width:7 ~clock [1; 2; 3]
\end{minted}

The constructed LFSR \verb|x| behaves identically to the following Verilog code
snippet:

\begin{minted}{verilog}
reg[6:0] x;
wire input_bit = x[1] ^ x[2] ^ x[3];
always @ (posedge clock) begin
  x <= { x[5:0], input_bit };
end
\end{minted}

\subsubsection{Constructing a RAM}

Hardcaml has built-in support for constructing memories, which are inferrable by backend
tools like Vivado. Models are provided for both simulation and synthesis, which have
been verified against vendor-provided RTL using the formal proving libraries provided with
Hardcaml and discussed in section 2.9.3.

\subsubsection{Constructing an ROM}

A ROM can be defined from an arbitrary OCaml math function as follows. This means that
Hardcaml is able to leverage the wide range of mathematical libraries available in the
OCaml ecosystem, for example \verb|math|, \verb|float|, and \verb|owl| (which provides sparse matrix and linear algebra functions).

\begin{minted}{ocaml}
let create_rom ~f ~clock ~size ~read_address =
  List.init size ~f
  |> mux read_address
  |> reg (Reg_spec.create ~clock ())

let create_sine_rom
    ~clock
    ~read_address
    ~scale
    ~size =
  create_rom
    ~clock
    ~f:(fun x -> of_int (Math.sin x * scale) ~width:32 )
    ~size
    ~read_address
\end{minted}

\subsection{Fixed Point Support}

Hardcaml also provides the \verb|hardcaml_fixed_point| library, with support for hardware-synthesizable
arithmetic operations such as \verb|+|, \verb|-|, and \verb|*|. It provides a rich set of rounding and
clamping modes, and seperate types for performing signed or unsigned operations. This
library is useful for many signal processing applications, where explicit control over
integer and fractional bit-widths is required. Functors can be used with
this library to implement and experiment different bit-widths and the resulting change of output precision.

\subsection{Testing and Simulation}

One of the aspects of Hardcaml that sets it apart from other DSLs is its emphasis
and support for circuit testing, simulation, and verification. The next sections highlight
the technologies that are provided with Hardcaml for this purpose.

\subsubsection{Expect-test Simulations and Waveforms}

The ability to simulate circuits, view resulting waveforms, and fix any bugs are important parts of the workflow.
While many other DSLs in this space require transformation through
an intermediate form and the use of external simulators and waveform display
tools~\cite{spinalhdl, chisel}, Hardcaml supports direct simulation of the
circuit graph inside OCaml. The cycle-based simulator has been optimized for runtime performance, and common primitive bit-level manipulation is done
via fast OCaml Foreign Function Interface (FFI) calls into C.

OCaml tests are written in a "snapshot" style, where statements can be
immediately followed by an \verb|[%expect]| block that prints that statement's
output. Any changes in the code that affect this frozen output are detected as
test failures. Hardcaml integrates into this flow seamlessly, and circuit inputs and outputs
can be printed at any cycle in the simulation.

\begin{figure}[!th]
\begin{minted}{ocaml}
module Circuit = Cyclesim.With_interface (Adder.I) (Adder.O)

let%expect_test "Adder works correctly" =
  let sim = Circuit.create ~config:Cyclesim.Config.trace_all Adder.create in
  let waves, sim = Waveform.create sim in
  let i = Cyclesim.inputs sim in
  i.inputs.(0) := Bits.of_int ~width:4 1;
  Cyclesim.cycle sim;
  i.inputs.(1) := Bits.of_int ~width:4 2;
  Cyclesim.cycle sim;
  Cyclesim.cycle sim;
  i.inputs.(0) := Bits.zero 4;
  i.inputs.(1) := Bits.zero 4;
  Cyclesim.cycle sim;
  Waveform.expect ~wave_width:1 ~display_width:50 ~display_height:17 waves;
\end{minted}
 \includegraphics[height=6cm]{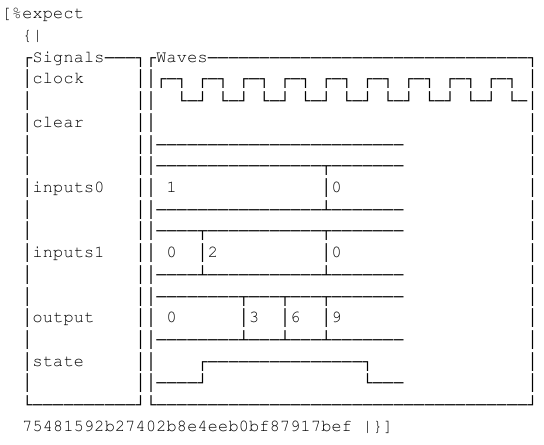}
\caption{Example Hardcaml simulation and ASCII waveform in an expect-test.}
\label{fig:waveform}
\end{figure}

Waveterm is one of Hardcaml's many supporting libraries. It can print ASCII waveforms over
a given number of cycles, and it can also operate as a stand-alone tool for interactive
use. It can also be integrated into project-specific testing binaries. The advantage of
ASCII waveforms is that they can be included in expect-test output. This helps users
understand how Hardcaml functions work. It also makes clear how changes in source code
change test outputs, which makes breaking changes easier to detect. This is also helpful
when reviewing changes submitted to a version-control repository, since the reviewer can
directly see how the code changes affect the output. This will also allow for quick
detection of any unwanted changes to source code that cause existing circuits tests to
fail. Figure \ref{fig:waveform} shows an example output of this type of test, on the adder
state-machine circuit previously introduced. Here we are assigning to our inputs with the
\verb|Bits.t| type, which is used to drive circuit inputs and can take concrete values.

\subsubsection{Integration into Other Simulation Backends}

In addition to the built-in simulator Cyclesim provided, Hardcaml also allows for seamless
simulation into two other backends - \verb|hardcaml_c| and \verb|hardcaml_verilator|
(Verilator version 5.0.14~\cite{verilator}). These provide better simulation performance at
the expense of a longer compilation time. \verb|hardcaml_c| uses even more bindings to C
for the circuit evaluation operations, and \verb|hardcaml_verilator| compiles the circuit
to a shared object and calls into a Verilator binary. When the simulation is complete, the
backend sends the results back to the Hardcaml framework, which displays them in the same
waveform viewers and debugging tools used with the built-in simulator. The user doesn't
need to change their workflow, other than calling a different simulator function. By
allowing seamless integration into these backends, simulation runtime can be optimized
depending on the size and number of input stimuli. Table \ref{table:simulator} shows a
comparison of the available backends. We created a large board-level network passing circuit in Hardcaml, which generated 180k lines of Verilog.
We measured simulation and compile times for 100, 1000, and 10,000 packets.

\begin{table}[!ht]
  \caption{Simulation and compilation times for various backends.}
  \begin{center}
{
    \begin{tabular}{|c|c|c|c|}
      \hline
       \multirow{2}{*}{\textbf{Packets}} & \multicolumn{3}{|c|}{Simulation time (s)} \\
       \cline{2-4}
       & \textbf{Cyclesim} & \textbf{Cyclesim\_c} & \textbf{Verilator} \\
      \hline
      100 & 6 & 8 & 2 \\
      \hline
      1,000 & 43 & 20 & 3 \\
      \hline
      10,000 & 419  & 138  & 17  \\
      \hline
      \hline
      Compilation time (s) & 1 & 8 & 263 \\
      \hline
    \end{tabular}}
    \label{table:simulator}
  \end{center}
\end{table}

\subsubsection{Hardcaml Verify}

When designing circuits where correctness is paramount, more detailed
verification is required. The library \verb|hardcaml_verify| provides many tools for the user:

\begin{itemize}
  \item SAT solvers can check the equivalence of combinational circuits and identify error cases.
  \item The Bounded Model Checker can check the equivalence of two sequential circuits, up to a given number of clock cycles.
  \item NuSMV~\cite{nusmv} integrates with Hardcaml to prove LTL assertions. While this is slow and works better for small circuits,
  Hardcaml makes it possible to parameterize the size of the circuit, so NuSMV can prove assertions for small versions of the circuit.
\end{itemize}

\subsubsection{Built-in Synthesis Reports}

Hardcaml also has a built-in synthesis report utility. This allows visualizing
hierarchical synthesis results of Hardcaml circuits by running them through backend tools.
For example, the programmer can setup a synthesis executable for the design below, which
instantiates multiple Hardcaml modules.

\begin{minted}{ocaml}
module Dut = struct
  module I = ...
  module O = ...

  let create scope { I.clock; a; b; c } =
    let spec = Reg_spec.create ~clock () in
    let adder1 =
      Adder_with_1_cycle_latency.hierarchical
        ~instance:"adder1"
        scope
        { clock; x = a; y = b }
    in
    let adder2 =
      Adder_with_1_cycle_latency.hierarchical
        ~instance:"adder2"
        scope
        { clock; x = adder1.z; y = reg spec c }
    in
    { O.d = adder2.x }
end

module Command = Synth.Command.With_interface(Dut.I)(Dut.O)

let () =
  Hardcaml_xilinx_reports.Comand.command_run
    ~name:"dut"
    Dut.create
\end{minted}

\begin{figure} [!t]
  \centerline{\includegraphics[width=\linewidth]{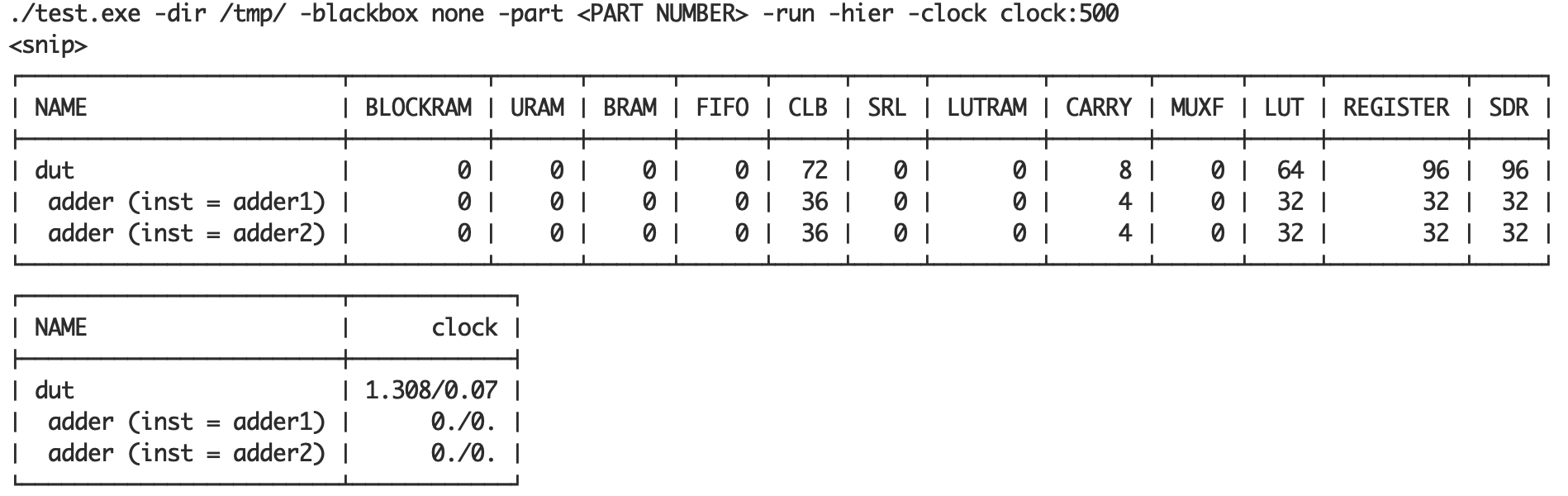}}
  \caption{Synthesis results output of our tool for the example design at 500MHz. The first table shows the hierarchical resource utilization. The second table shows the hierarchical timing slack estimated from synthesis.}
  \label{fig:example_synthesis_results}
\end{figure}

The user can then run the executable to obtain a table with the timing summary and
resource utilization of the synthesis results, as show in an example in figure
\ref{fig:example_synthesis_results}. The ease of running this tool lowers the barriers to
running synthesis on small circuits. There are also more options available to control the
output artifacts, such as running place and route on top of synthesis, or generating
intermediate design checkpoint files.

\subsection{Drawbacks of Hardcaml}

Although Hardcaml has been able to build upon the advantages other DSLs offer, there are still several shortcomings when compared to Verilog and VHDL.
\begin{itemize}
\item Unless specifically labeled, the mapping of Hardcaml function calls to their generated RTL might not be obvious to a user, which can be important when debugging timing violations in Vivado.
\item Interfacing with externally written IP cores requires blackbox wrappers in Hardcaml which can't be simulated, without doing additional work such as importing the RTL into Hardcaml.
\item The user may not be familiar with functional programming or OCaml.
\end{itemize}

\subsection{Future Work}

We are continuously adding new features to Hardcaml and releasing improvements into open
source repositories. We plan to implement the following new features that will further
differentiate Hardcaml from other DSLs:

\paragraph{Encoding Signal Clock-Domains within the Type System}

Currently, the entire design of a Hardcaml circuit uses the same \verb|Signal.t| type. A
more advanced approach is to constrain \verb|Signal.t| to a particular clock domain using
a functional programming technique called phantom types. This uses the type system to
ensure that the programmer did not accidentally mix signals between different clock
domains. When cross-clock-domain crossing is necessary, signals can be marked as such and
this validation will be relaxed. This can provide a balance between clock-domain-safety
and flexibility in real-world designs.

\paragraph{Floating Point Support}

We are implementing a suite of floating point cores, including adders, multipliers,
fixed-point converters, and more. We will provide these in the open-source Hardcaml suite
when they are ready.

\paragraph{Integration with Higher-Level HDLs}

As mentioned in the introduction, some HDLs like Filament~\cite{filament} and
Bluespec~\cite{bluespec} trade-off low-level control for a higher level of
abstraction. A seamless integration (eg: within a Hardcaml library) between
Hardcaml circuits with low-level control and these HDLs can potentially further
improve programmer productivity by giving them more tools in the Hardcaml
development environment. This will include features like a
user-friendly API to construct circuits in those HDLs, rapid
circuit-elaboration-time validation, integration with Hardcaml's existing
simulation frameworks, and RTL generation.

\section{A Case Study into Using Hardcaml}

%CR bdevlin: Put this text back
%In 2022 there was a competition to design a cryptographic FPGA accelerator - the
%ZPrize\cite{zprize}. We submitted a Multi-Scalar Multiplication (MSM) engine written
%entirely in Hardcaml, Hardcaml MSM\cite{hardcaml-msm}, and won 1st place.
We designed a Multi-Scalar Multiplication (MSM) engine written
entirely in Hardcaml, Hardcaml MSM.
In this section we compare and highlight several of the design techniques from Hardcaml that enabled us to
efficiently design this engine with a tight deadline and limited human resources.

\subsection{Config Functorization}

We used the \verb|Config| module, shown below, to parameterize our design. We could easily
change bit widths, window bounds, and circuit partitioning over the 3 Super Logic Regions
(SLRs) available to us on the FPGA. Within seconds of making a change, the Dune build
system would tell us if any of our tests failed, or if our design's performance changed.
In Verilog, we would not have been able to parameterize our design to this degree, and
iterating over the design space would have taken much longer.

\begin{minted}{ocaml}
type slr =
  | SLR0
  | SLR1
  | SLR2
[@@deriving sexp_of]

type partition_setting =
  { num_windows : int
  ; slr : slr
  }
[@@deriving sexp_of]

module Config : sig
  val field_bits : int
  val scalar_bits : int
  val controller_log_stall_fifo_depth : int
  val num_windows : int
  val window_ram_partition_settings : partition_setting list option
end

\end{minted}

%% CR bdevlin: Add examples of the Config with different values?

\subsection{Adding Register Stages}

Inserting registers to pipeline a design is a common method to improve
throughput. Doing this in Hardcaml was very easy due to the auto-generated
utility functions on our custom types. The following code creates a new module
\verb|foo|, which uses a pipelined version of the AXI stream \verb|my_stream|.
We create the pipelined stream using Stream.Source.map, which creates a register
for each signal in \verb|my_stream| and registers all downstream connections.

\begin{minted}{ocaml}
let foo =
  Foo.create
    { my_stream =
        Stream.Source.map my_stream ~f:(reg spec)
    }
in
\end{minted}

In Verilog this would involve multiple registers for the different signals in the bus
such as \verb|tlast|, \verb|tvalid|, \verb|tdata|, an always block, and
correct assignment to the downstream block.

\begin{minted}{verilog}
reg my_stream_tvalid_r;
reg my_stream_tlast_r;
reg my_stream_tdata_r;

always @(posedge clock) begin
  my_stream_tvalid_r <= my_stream_tvalid;
  my_stream_tlast_r  <= my_stream_tlast;
  my_stream_tdata_r  <= my_stream_tdata;
end

foo foo (
  .tvalid(my_stream_tvalid_r),
  .tdata(my_stream_tdata_r),
  .tlast(my_stream_tlast_r)
);
\end{minted}

\subsection{Breaking Computation into Pipelined Stages}

Hardcaml MSM contains a multi-stage pipelined elliptic curve point adder.
Hardcaml allows us to easily break the computation into several stages. A
simplified form of this pattern looks something like the following:

\begin{minted}{ocaml}
module Stage0 = struct
  type 'a t =
    { x : 'a [@bits 16]
    }
  [@@deriving hardcaml]
end

module Stage1 = struct
  type 'a t =
    { y : 'a [@bits 32]
    }
  [@@deriving hardcaml]

  let create { Stage0. x } = { y = x *: x }
end

module Stage2 = struct
  type 'a t =
    { z : 'a [@bits 32]
    }
  [@@deriving hardcaml]

  let create { Stage1. y } = { z = y +:. 1 }
end

let pipelined_computation ~clock ~x =
  let spec = Reg_spec.create ~clock () in
  let result =
     { Stage0. x }
     |> Stage1.create
     |> Stage1.map ~f:(reg spec)
     |> Stage2.create
     |> Stage2.map ~f:(reg spec)
   in
   result.y
\end{minted}

This design pattern has two major benefits:
\begin{enumerate}
\item  Within a pipeline stage, all the signals are combinational, which avoids using a signal
from a different pipeline stage that represents a different calculation
\item Different pipeline stages can easily be tested in isolation
\end{enumerate}

\subsection{Availability of Model Libraries}

This project required a lot of modulo arithmetic on very large numbers. We used existing
OCaml libraries \verb|zarith| and \verb|bigint|, which worked with Hardcaml even though they were
written for OCaml software programs.

\subsubsection{Computing Complicated Functions on the Fly}

Hardcaml can compute complicated elaboration-time functions over design constants without
hardcoding these as parameters. This allows us to limit the set of parameters of a module
to those that are truly necessary. For example, a module in this project was parameterized
over a large constant $c$. We needed both $c$ and $c^{-1} \bmod P$ in our circuit. In
Verilog, we would have had to provide both $c$ and $c^{-1}$ as parameters, but in
Hardcaml, we could pass in $c$ and compute $c^{-1}$ at elaboration time.

\subsubsection{Testing against Model Libraries}

Because these libraries are available for OCaml as well as Hardcaml, we could write a
simple software implementation of parts of our circuit. Then, we could provide the same
randomized inputs to both implementations. We used expect tests to compare their outputs,
making it easy to write hand-crafted test cases. In Verilog, we would have needed to put
our inputs in a static file, provide them to the Verilog simulation, separately provide
them to a software implementation, and then use another tool to compare the outputs.

\section{Conclusion}

This paper introduces Hardcaml - an embedded hardware design domain
specific language (DSL) implemented in the OCaml programming language. Unlike high level
synthesis (HLS), designing in Hardcaml allows for low level control of the underlying
hardware for maximum productivity and performance, while abstracting away much of the
tedious aspects present when designing in traditional HDL languages such as Verilog, System Verilog,
or VHDL. Robust design comes by using the OCaml type system to encode important
properties of hardware primitives such that connecting modules and transformations on
signals can be programmatically applied in a software-like manner, reducing the chance of
user-introduced bugs and erroneous connections. Hardcaml includes both a native OCaml
cycle-accurate and event-driven simulator, along with backend support for external simulators such as Cyclesim\_c and
Verilator. This allows for unit level tests alongside the Hardcaml source
code, which can optionally print digital ASCII waveforms. For several years Hardcaml has
also been industrially proven in many FPGA designs internally in Jane Street and has
proven indispensable in the benefits it has provided.
As a case study we also highlight
several of the benefits of Hardcaml in our submission to the 2022 ZPrize, which won 1st
place in the FPGA track.

\section*{Acknowledgment}

We would like to thank Pranjal Vachaspati for constructive criticism and proofreading of the paper.

\bibliographystyle{ACM-Reference-Format}
\bibliography{bibliography}

\end{document}